\documentclass[12pt]{article}

\usepackage{epsf}
\usepackage{latexsym,euscript}
\usepackage[dvips]{graphicx}
\usepackage{amsmath}
\usepackage{amsfonts}
\usepackage{amssymb, epsfig}
\usepackage{cite}

\usepackage{color,soul}

\begin{document}

\begin{center}
{\Large \textbf{The Polyakov loop models in the large $N$ limit: \\
Correlation function and screening masses}}

\vspace*{0.6cm}
\textbf{O.~Borisenko${}^{\rm a,b}$\footnote{email: oleg@bitp.kiev.ua},
V.~Chelnokov${}^{\rm c}$\footnote{email: chelnokov@itp.uni-frankfurt.de},
S.~Voloshyn${}^{\rm b}$\footnote{email: s.voloshyn@bitp.kiev.ua}}

\vspace*{0.3cm}
{\large \textit{${}^{\rm a}$ INFN Gruppo Collegato di Cosenza, Arcavacata di Rende, 87036 Cosenza, Italy}} \\
{\large \textit{${}^{\rm b}$ Bogolyubov Institute for Theoretical
Physics, National Academy of Sciences of Ukraine, 03143 Kyiv, Ukraine}} \\
{\large \textit{${}^{\rm c}$ Institut f\"ur Theoretische Physik, Goethe-Universit\"at Frankfurt, 60438 Frankfurt am Main, Germany}}
\end{center}

\begin{abstract}
We  explore the  't Hooft-Veneziano limit of the Polyakov loop models at finite baryon chemical potential. Using methods developed by us earlier we calculate the two- and $N$-point correlation functions of the Polyakov loops. This gives a possibility to compute the various potentials in the confinement phase and to derive the screening masses outside the confinement region.
In particular, we establish the existence of complex masses
and an oscillating decay of correlations in a certain range of parameters.
Furthermore, it is shown that the calculation of the $N$-point correlation function in the confinement phase reduces to the geometric median problem. This leads to a large $N$ analog of the $Y$ law for the baryon potential.
\end{abstract}

\section{Introduction}

The characteristics of strongly interacting matter
at high temperatures and densities are actively studied
through theoretical and computational methods.
The phase diagram of Quantum Chromodynamics (QCD),
which shows how strongly interacting matter behaves
as temperature and baryon density change,
is currently a subject of intense investigation.
Studying the QCD phase diagram is crucial in gaining insights into the
fundamental properties of matter under extreme conditions,
such as those one encounters in heavy ion collisions
or within compact astrophysical objects.

It is well known that the introduction of
a chemical potential $\mu$ into the QCD action
makes the Euclidean path integral measure complex,
so standard Monte-Carlo simulations are not feasible.
Over the last few decades, various approaches have been developed
to tackle this sign problem either partially or entirely.
Some notable methods include Taylor expansion or reweighting
at low chemical potential, simulations at imaginary potential,
complex Langevin simulations,
and others as reviewed in \cite{philipsen_rev_19,Seiler}.
Despite certain progress achieved within these methods, lattice simulations
at arbitrary real chemical potential are not yet possible.

Analytical efforts to overcome the sign problem are mainly concentrated around
some effective theories, like the Polyakov loop models, with the goal to map such theories via the duality transformations to models with the positive Boltzmann weight, see \cite{Gattringer11,Gattringer12a,Philipsen12,un_dual18,pl_dual20}.
Another direction of analytical attempts is investigation of the 't Hooft and the
't Hooft-Veneziano limits of lattice QCD. {\it E.g.}, using methods developed in
\cite{gross_witten,wadia}, the large $N$ limit was explored for $U(N)$ Polyakov loop models in \cite{damgaard_patkos,christensen12}.
Extension of  these methods to $SU(N)$ models was accomplished in
Refs.~\cite{largeN_sun,pl_largeN_conf21,pl_largeN21}. In these papers we established a general phase structure of the model in the 't Hooft-Veneziano limit. 
An important open problem is the behavior of the Polyakov loop correlations in this limit. This problem is addressed in the present paper.
The large distance decay of the Polyakov loop correlations at finite chemical potential is governed by the (electric and magnetic) screening masses.
For a general review on screening masses, we refer to \cite{Bazavov}.
In Refs.~\cite{Nagata13,Andreoli} these masses have been computed
in lattice QCD with imaginary chemical potential.
Relatively little is known about screening masses in the presence of the real chemical potential. Some results obtained from the simulations of the dual of the $SU(3)$ Polyakov loop model can be found in Refs.~\cite{mcdual_21,su3_dual_cooreleations}.

A closely related problem is the emergence of the complex spectrum at finite density. This leads to an exponential decay of the correlations modulated by an oscillating function \cite{Meisner2011,ogilvie2016,forcrand_z3,duals_abelian}.
Locating such a liquid phase requires the computation of long-distance correlations
with real chemical potential. So far, a phase with an oscillating decay of correlations was shown to exist in $(1+1)d$ LGT with heavy quarks in \cite{ogilvie2016} and in $Z(3)$ spin model in a complex external field in \cite{forcrand_z3}. The latter result was extended to many $Z(N>3)$ models in
\cite{duals_abelian}.
Here we prove that the liquid phase exists in the 't Hooft-Veneziano limit
of the Polyakov loop model in a certain range of parameters. Preliminary results
of this study have been presented in \cite{pl_largeN_conf21}.

We work with the Polyakov loop model whose action is given by
\begin{eqnarray}
S  =  \beta \ \sum_{x,n} \ {\rm Re} W(x) W^*(x+e_{n}) +
\sum_x \sum_{f=1}^{N_f} h(m_f)
\left ( e^{\mu} W(x) + e^{-\mu} W^*(x) \right ) \ .
\label{action_local}
\end{eqnarray}
Here, $W(x)= {\rm Tr} U(x), U(x)\in U(N), SU(N)$,
$\beta=\beta(g^2, N_t)$ is an effective coupling constant,
$h(m_f)$ is a function of the quark mass $m_f$ and $N_f$ is a number of fermion flavors. Our goal is to calculate the correlation functions
of this model in the 't Hooft-Veneziano limit \cite{Hooft_74,Veneziano_76}:
$g\to 0, N\to\infty, N_f\to\infty$ such that the product $g^2 N$
and the ratio $N_f/N=\kappa$ are kept fixed.
For the case of $N_f$ degenerate flavors considered here, one has:
$\sum_{f=1}^{N_f} h(m_f) = N_f h(m) \to N \kappa h(m) \equiv N \alpha$.
%\begin{equation}
%$\sum_{f=1}^{N_f} h(m_f) = N_f h(m) \to N \kappa h(m) \equiv N \alpha$ \ .
%	\label{alpha_def}
%\end{equation}
In the previous  papers \cite{pl_largeN_conf21,pl_largeN21}
we derived the large $N$ representation of the model and described its phase diagram. 
In particular, a third order phase transition separating two phases has been found. 
Let us briefly describe the large $N$ representation which is the starting point of the following calculations. 
If $\eta(x)$ ($\bar{\eta}(x)$) is the power of the Polyakov loop (its conjugate) then the arbitrary correlation function can be written down as
\begin{eqnarray}
\Gamma =  e^{\mu \sum_x(\bar{\eta}(x) - \eta(x))} \ \left \langle \prod_x \rho(x)^{\eta(x)+\bar{\eta}(x)} \ e^{i\omega(x)(\eta(x)-\bar{\eta}(x))}
 \right \rangle \  .
\label{corr_1}
\end{eqnarray}
The expectation value in the last expression refers to the following
partition function
\begin{equation}
Z = \prod_x \ \int_0^1\rho(x)d\rho(x) \int_0^{2\pi} \frac{d\omega(x)}{2\pi}
\int_{-\infty}^{\infty} \ du(x) dt(x) ds(x) \ e^{N^2 S_{eff}} \ .
\label{PF_largeN}
\end{equation}
The effective action reads
\begin{eqnarray}
S_{eff} &=& \beta \ \sum_{x,n} \ \rho(x) \rho(x+e_n) \cos\left ( \omega(x) - \omega(x+e_n) \right )
+ \alpha \ \sum_x \rho(x) \cos\omega(x) \nonumber  \\
\label{Seff_1}
&+& \mu \sum_x u(x) + \sum_x \  V(\rho(x),\omega(x);u(x),t(x),s(x))  \ , \\
V(\rho,\omega;u,t,s) &=& -iu\omega + \frac{3}{2} |u| - \frac{1}{2} (1+|u|)^2\ln (1+|u|) + \frac{1}{2} u^2 \ln |u| -2i\rho s  \nonumber  \\
&+& |u|\ln (t+is) - t^2 -s^2 - \sum_{k=0}^{\infty} (t^2+s^2)^{k+1} C_k(|u|) \ .
\nonumber
\end{eqnarray}
The function $C_k(|u|)$ is given in Appendix.
The  model exhibits the phase transition of third order at $z=0$,
where $z$ is defined as
\begin{equation}
z =	\mu - \ln \left ( 1 + \sqrt{1 - \frac{\alpha^2}{(1- \beta d)^2}} \right ) + \ln \frac{\alpha}{1- \beta d} + \sqrt{1 - \frac{\alpha^2}{(1- \beta d)^2}}  \ .
	\label{crit_line_sun}
\end{equation}

%The rest of the paper is organized as follows. In Sec.2  we calculate the general %form of correlation functions and study the behavior of the string tension and %screening masses. Complex masses appearing above the phase transition are %described in details. In Sec.3 the $N$-point function related to baryon potential %is evaluated in the confinement phase of the pure gauge theory.

\section{Correlation functions and screening masses}

In this section we calculate various correlation functions
and corresponding screening masses in the large $N$ limit.
As is well known, at nonzero chemical potential the Polyakov loop correlations
form a correlation matrix \cite{Andreoli}
\begin{eqnarray}
\Gamma (x,y)\!  = \! \begin{pmatrix}
\left \langle  \frac{1}{N}\mbox{Re} W(x)    \,  \frac{1}{N}\mbox{Re} W^{\dagger}(y)  \right \rangle_c   & \! \! \! \! \! \left \langle  \frac{1}{N}\mbox{Re} W(x) \, \frac{1}{N}\mbox{Im} W^{\dagger}(y)  \right \rangle_c \\
\left \langle  \frac{1}{N}\mbox{Im} W(x) \, \frac{1}{N}\mbox{Re} W^{\dagger}(y)  \right \rangle_c &  \! \! \! \! \! \left \langle   \frac{1}{N}\mbox{Im} W(x) \, \frac{1}{N}\mbox{Im} W^{\dagger}(y)  \right \rangle_c
\end{pmatrix} \! = \!
\begin{pmatrix}
 \Gamma_{rr}   & \Gamma_{ri} \\
\Gamma_{ir} & \Gamma_{ii}
\end{pmatrix}\ ,
\label{corr_matrix}
\end{eqnarray}
where $\langle\ldots\rangle_c$ refers to a connected part of the correlation.
When $\mu = 0$ the off-diagonal terms vanish and the coefficients in the exponential decay of diagonal terms define the magnetic and electric screening masses
\begin{equation}
\label{masses_screen_def}
\Gamma_{rr} \simeq \frac{e^{- m_M R}}{R^{\eta}} \ \ , \ \Gamma_{ii} \simeq \frac{e^{- m_E R}}{R^{\eta}} \ , \  R=|x-y| \ .
\end{equation}
When $\mu> 0$ the electric and magnetic sectors mix, so each correlation matrix element is a sum of two terms -- one decaying with $m_M$ and the other with $m_E$.

In the limit $N\to\infty$ all integrals in (\ref{PF_largeN}) are evaluated by the saddle-point method. 
In what follows we denote by $\rho_0,\omega_0,u_0,t_0,s_0$ the corresponding saddle points. 
For $SU(N)$ model the full analytical solution can be obtained near the critical surface $z=0$. 
For completeness, we give this solution in Appendix.
In the large $N$ limit, the expectation value of the Polyakov loop equals
$\rho_0 e^{\pm i\omega_0}$ and the invariant two-point correlation function
is simply $\rho_0^2$. Therefore, nontrivial correlations appear only in sub-leading
terms that can be computed by expanding the full action around saddle points, {\it e.g.} $\rho(x) \to \rho_0 + \frac{1}{N} \rho(x)$ and so on,
and by evaluating the resulting Gaussian integrals over fluctuations.
Since all calculations are straightforward we present and discuss below the final results for the correlations.  
We use the following notation  for the Green function appearing in the results
\begin{equation}
G_{x,x^{\prime}}(m) = \frac{1}{L^d} \ \sum_{k_n=0}^{L-1} \
\frac{e^{\frac{2\pi i}{L} \sum_n^d k_n (x_n-x_n^{\prime})}}{m + f(k)} \ , \ f(k) = d-\sum_{n=1}^d\cos\frac{2\pi}{L}k_n \ .
\label{Green_func}
\end{equation}

\subsection{Two-point functions in $U(N)$ model}

For $U(N)$ model, following \cite{pl_largeN21}, one finds $u_0=t_0=\omega_0=0$,
$s_0=-i \left ( \frac{\alpha}{2} + \beta d\rho_0 \right )$ and
\begin{eqnarray}
	\rho_0 \ = \
	\begin{cases}
\frac{\alpha}{2(1-\beta d)}  \ , \    \ \ \ \alpha + \beta d \leq 1  \ , \\
\frac{1}{4\beta d} \ \left ( 2\beta d -\alpha + \sqrt{(2\beta d + \alpha)^2-4\beta d}   \right )  \ , \ \ \ \alpha + \beta d \geq 1  \ .
	\end{cases}
	\label{rho_0_un}
\end{eqnarray}
In the pure gauge case, $\alpha=0$, one observes a first order
confinement-deconfinement phase transition.
The expectation value of the Polyakov loop jumps from zero to $1/2$
at the critical point. When $\alpha$ is nonzero, the system undergoes
the third order phase transition. Accordingly, we describe two-point correlations for these cases.
\begin{enumerate}
\item
$\alpha=0$, confinement phase:
\begin{equation}
\Gamma_2(R) = \langle \ W(0) W^*(R) \ \rangle =
\frac{1}{\beta N^2} \ G_R(m_1) \ , \ m_1 = \frac{1}{\beta} - d \ .
\label{gamma2_un_1}
\end{equation}

\item
$\alpha=0$, deconfinement phase:
\begin{eqnarray}
\Gamma_2(R) &=& \rho_0^2 \exp\left [ \frac{1}{2\beta\rho_0^2 N^2} (G_0(m_2)+G_R(m_2) - G_0(0) + G_R(0))  \right] \ , \nonumber  \\
m_2 &=& \frac{1}{\beta} \left ( 1- \sqrt{1-1/\beta d} \right )^{-2} - d \ .
\label{gamma2_un_2}
\end{eqnarray}

\item
$\alpha\ne 0$, confinement phase:
\begin{eqnarray}
\Gamma_2(R) &=& \rho_0^2 \
\exp\left [\frac{1}{\beta\rho_0^2 N^2} \ G_R(m_1) \right] \  .
\label{gamma2_un_3}
\end{eqnarray}

\item
$\alpha\ne 0$, deconfinement phase:
\begin{eqnarray}
\Gamma_2(R) &=& \rho_0^2 \exp\left [ \frac{1}{2\beta\rho_0^2 N^2} (G_0(m_3)+G_R(m_3) - G_0(m_4) + G_R(m_4))  \right] \ , \nonumber  \\
m_3 &=& \frac{1}{4 \beta (1-\rho_0)^2}  - d \ , \ \  m_4 \ = \ \frac{ \alpha  }{2 \beta \rho_0} \ ,
\label{gamma2_un_4}
\end{eqnarray}
where $\rho_0$ is given in the bottom line of (\ref{rho_0_un}).

\end{enumerate}
The dependence on the chemical potential drops out in $U(N)$ model both from the free energy and from invariant observables. Screening masses do not depend on $\mu$.

\subsection{Correlation functions in $SU(N)$ model}

For $SU(N)$ model  we obtain the following general result when $z>0$
\begin{align}
\label{correlator_res}
&\Gamma = \prod_x e^{(\mu - i\omega_0)(\bar{\eta}(x)-\eta(x))} \
\rho_0^{(\eta(x) + \bar{\eta}(x))}
\exp\left [ \frac{1}{4 N^2 \beta \rho_0^2}  \sum_{x,x^{\prime}}
\left (  A_1(x,x^{\prime}) + A_2(x,x^{\prime}) \right ) \right ]  \\
\label{A1_notation}
&A_1(x,x^{\prime}) = \frac{1}{\sqrt{C_1 C_2}} \ \left ( C_1 \eta(x)\eta(x^{\prime})
+ C_2 \bar{\eta}(x)\bar{\eta}(x^{\prime})  \right ) \
\left ( G_{x,x^{\prime}}(m_{-}) - G_{x,x^{\prime}}(m_{+}) \right )  \ , \\
\label{A2_notation}
&A_2(x,x^{\prime}) = 2 \eta(x) \bar{\eta}(x^{\prime}) \
\left ( G_{x,x^{\prime}}(m_{-}) + G_{x,x^{\prime}}(m_{+}) \right )  \ .
\end{align}
The masses $m_+$ and $m_-$ are given by
\begin{equation}
	m_\pm \ = \ \frac{1}{2 \beta \rho_0^2} \
	\left ( C_3 \pm \sqrt{C_1 C_2}  \right ) \ ,
	\label{masse_def}
\end{equation}
where the constants $C_i$ are defined as
\begin{eqnarray}
	\label{C12_def}
	C_{1,2} =  b_2 + (d \beta  - b_1) \rho_0^2 \pm i b_3 \rho_0 \ ,  \
	C_3 = b_2 - (d \beta - b_1) \rho_0^2 \ .
%	\label{C3_def}
\end{eqnarray}
Coefficients $b_i$ are given in Appendix, Eqs.~(\ref{B_1})--(\ref{B_3}).
This result allows one to compute any observable above the critical surface $z>0$ by choosing the appropriate values of the sources $\eta(x)$ and $\bar{\eta}(x)$.  
The magnetization $M=\langle \rho(x) e^{i\omega(x)} \rangle$ becomes
\begin{eqnarray} 
\label{magnetization} 
M = \rho_0 e^{i\omega_0} \ \exp\left [ \frac{C_1}{4 N^2 \beta \rho_0^2 \sqrt{C_1 C_2}} \ \left (   G_0(m_{-}) - G_0(m_{+}) \right ) \right ]  \ ,  
\end{eqnarray}
where $G_0(m)$ is the zero distance Green function.
Noticing that $m_{-}\leq m_{+}$ one obtains for the eigenvalues of the correlation matrix (\ref{corr_matrix}) in the limit of a large separation $R$
\begin{eqnarray}
\label{M1}
{\cal{M}}_{1} &=& \frac{M M^*}{2N^2\beta \rho_0^2} \ G_R(m_{-}) \ , \\
\label{M2}
{\cal{M}}_{2} &=&   
  \begin{cases} 
    \frac{M M^*}{2N^2\beta \rho_0^2} \ G_R(m_{+}) \ ,  \mbox{if} \ m_{-} \leq m_{+} \leq 2 m_{-} \ ,  \\  
    \frac{M M^*}{2N^2\beta \rho_0^2} \left (  \ G_R(m_{+}) + \frac{1}{16 N^2\beta\rho_0^2}   \left ( 2-\frac{C_1+C_2}{\sqrt{C_1 C_2}} \right ) \ G_R^2(m_{-}) \right )  \ , \  \mbox{if} \ m_{+} \geq 2 m_{-} \ .  
  \end{cases}
\end{eqnarray}
Elements of the correlation matrix are found to be
\begin{eqnarray}
\label{real_real_corr}
\Gamma_{rr} = D \left [ ( \sqrt{C_1} M + \sqrt{C_2} M^*)^2 G_R(m_{-}) - ( \sqrt{C_1} M - \sqrt{C_2} M^*)^2 G_R(m_{+}) \right ] \ ,  \\
\label{im_im_corr}
\Gamma_{ii} = D \left [ ( \sqrt{C_1} M + \sqrt{C_2} M^*)^2 G_R(m_{+}) - ( \sqrt{C_1} M - \sqrt{C_2} M^*)^2 G_R(m_{-}) \right ] \ , \\
\label{real_im_corr}
\Gamma_{ri} =  D \left [ ( C_1 M^2 - C_2 M^{*,2}) (G_R(m_{-}) - G_R(m_{+}))  \right ]
\ , \  D = \frac{1}{8 N^2 \beta \rho_0^2 \sqrt{C_1 C_2}} \ .
\end{eqnarray}

\begin{figure}[h!]
%\centering{ \includegraphics[scale=0.6]{phaseD01nnn.pdf} \ \ \ %\includegraphics[scale=0.62]{phaseD05b.pdf}}
\centering{ \includegraphics[scale=0.69]{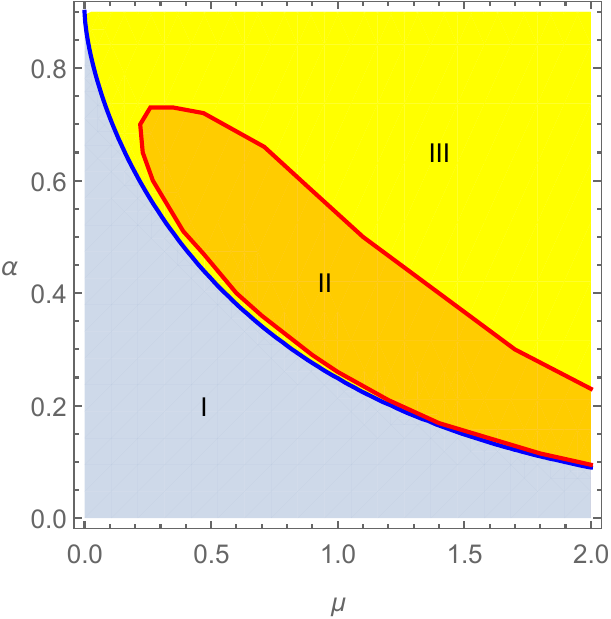} \ \ \ 
\includegraphics[scale=0.7]{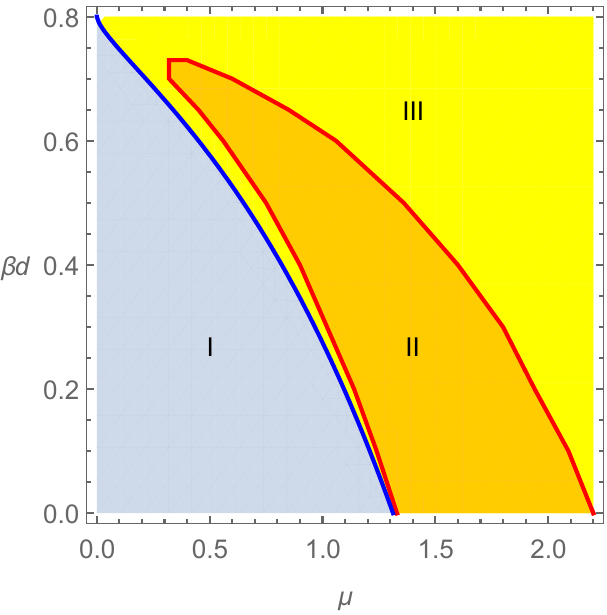}}
\caption{Cross-sections of the phase diagram of the SU(N) model. 
Left: $\mu$--$\alpha$ coordinates at fixed $\beta d=0.1$. 
Right: $\mu$--$\beta d$ coordinates at fixed $\alpha =0.2$. 
The blue curve shows the critical line of the 3rd order phase transition. See text for details.}
\label{fig1:phase_diagram12}
\end{figure}

\subsection{Phase diagram and complex masses}
\label{compl_mass}

The general phase structure of the $SU(N)$ model has been described in \cite{pl_largeN21}. Fig.~\ref{fig1:phase_diagram12} shows cross-sections of the phase diagram for a fixed value of $\beta d$ (left panel) and a fixed value of $\alpha$ (left panel), $d=3$. These cross-sections reveal three regions of the phase diagram which can be characterized by a different behavior of correlation functions and screening masses.

Region I. Here, one finds a single mass equal to the mass in $U(N)$ model for the same values of parameters. The free energy and correlation functions do not depend on the chemical potential. Crossing the transition line (blue line in Fig.~\ref{fig1:phase_diagram12}) one enters the region III which becomes more and more narrow with $\mu$ increasing. 
Analytical expressions for the masses can be found in the vicinity of
this transition line using explicit formulas for coefficients $b_i$
and saddle points given in Appendix. One finds
\begin{eqnarray}
m_{+,-}= \frac{1}{\beta} - d -\frac{2}{\beta  \left(4+\ln \frac{ z^2}{1- \frac{\alpha ^2}{( \beta d-1)^2}} \right)} \mp \frac{2}{\beta  \left(4+\ln \frac{ z^2}{1- \frac{\alpha ^2}{( \beta d-1)^2}} \right)} + {\cal {O}}(z) \ .
\label{m12_expan}
\end{eqnarray}

Region II. Crossing the lower red line in Fig.~\ref{fig1:phase_diagram12} the system moves to a phase, where a complex spectrum emerges. Masses $m_+$ and $m_-$ become conjugate to each other: $m_+=m_-^*=m_r+i m_i$.
The connected part of the two-point correlation is
\begin{equation}
\Gamma_2(R) \approx  M M^* (G_R(m_+) + G_R(m_-))
\sim \ \frac{e^{-m_r R}}{R} \ \cos m_i R  \ ,
\label{2point_corr}
\end{equation}
{\it i.e.} it has an exponential decay modulated by the cosine function.

\begin{figure}[htb]
\centering{ \includegraphics[scale=0.65]{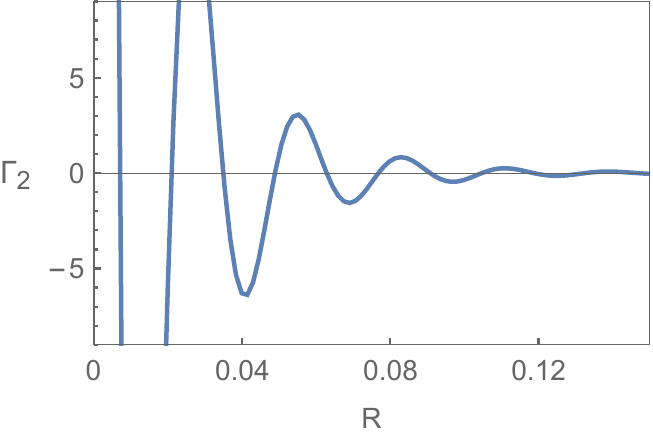} \ \ \ \includegraphics[scale=0.65]{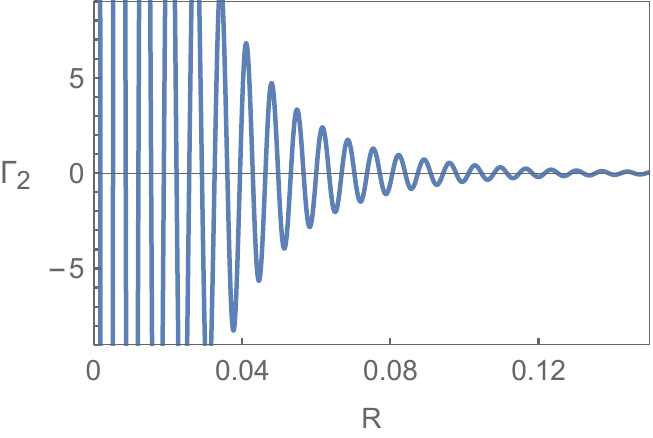}}
\caption{Oscillating decay of the correlation function with a distance in the Region II of the phase diagram. Left: $m_i/m_r \approx 7$, $\alpha=0.05$, $\beta d=0.1$, $\mu=2.595$. Right: $m_i/m_r \approx 30$, $\alpha=0.01$, $\beta d=0.1$, $\mu=4.1935$.}
\label{fig1:corrfunc}
\end{figure}

\begin{figure}[h!]
\centering{\ \includegraphics[scale=0.66]{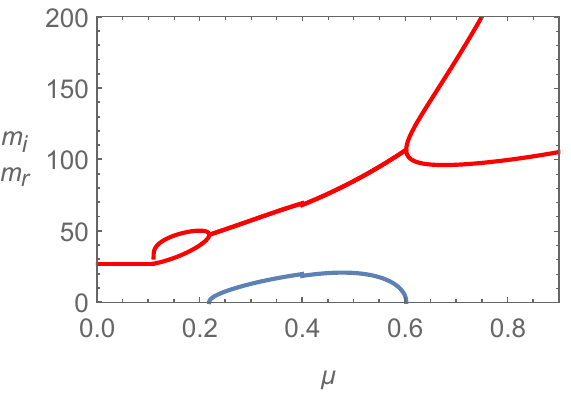} \ \ \  \includegraphics[scale=0.66]{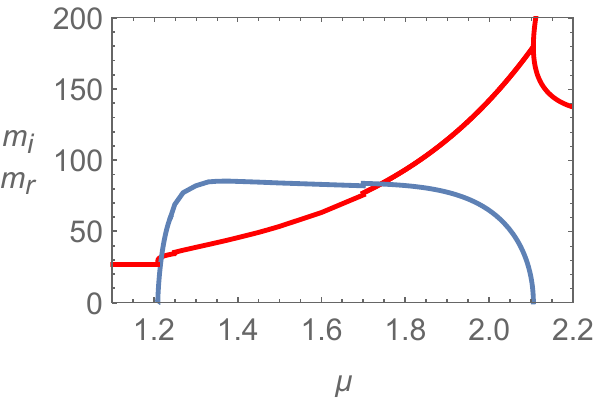} } \centering{ \includegraphics[scale=0.66]{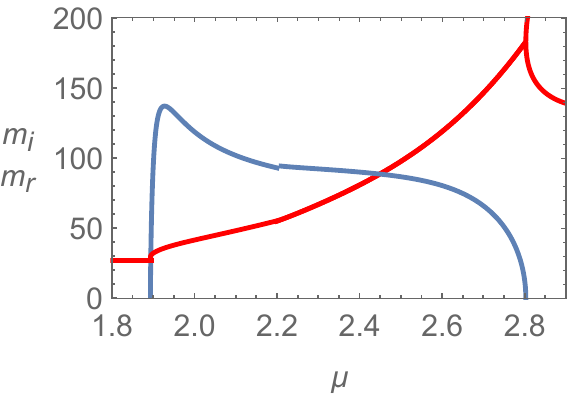} \ \ \	\includegraphics[scale=0.66]{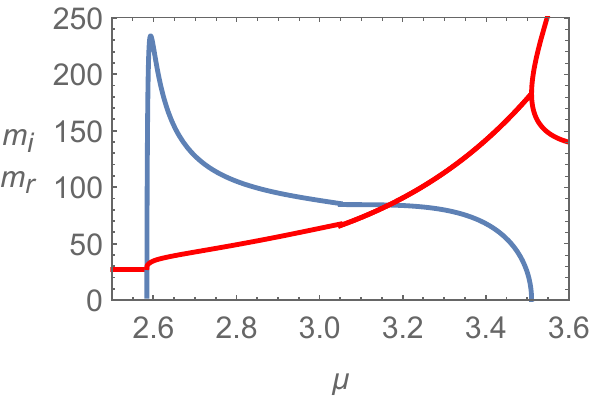}}
\caption{Real (red) and imaginary (blue) parts of the screening masses vs. $\mu$ at fixed $\beta d=0.1$ and fixed $\alpha=0.7$ (top left), $\alpha=0.2$ (top right), $\alpha=0.1$ (bottom left), $\alpha=0.05$ (bottom right). See text for details.}
\label{fig1:mass_ratio_4}
\end{figure}

When $\rho < 0.41$ and $\mu$ is sufficiently large one has $m_{i}>m_{r}$.
A maximum of the ratio $m_{i}/m_{r} $ is reached close to the phase transition.
The smaller $\alpha$ and the larger $\beta$ are, the more profound oscillations are observed. The corresponding behavior of the two-point correlation function is reflected in Fig.~\ref{fig1:corrfunc}.

Region III, separated from region II by a red line on Fig.~\ref{fig1:phase_diagram12}, has two distinct real masses and correlations decay exponentially according to Eqs.~(\ref{real_real_corr})-(\ref{real_im_corr}).
The equation defining the red line reads $C_2=0$.
This line does not define a genuine phase transition, since
the partition function together with all its derivatives remains analytical
in the thermodynamic limit. 

Finally, Fig.~\ref{fig1:mass_ratio_4} shows the screening masses as a function
of $\mu$ for $\beta d=0.1$ and several values of $\alpha$.
Two real masses above the 3rd order phase transition can be only seen on the top left panel (red line). When $\alpha$ is decreasing the region III gets narrower and  two masses become hardly distinguishable. Values of $\mu$, where the imaginary part is not zero, define the region II of the phase diagram.
A maximum of the imaginary part is located close to the smallest $\mu$ value.
In the upper region III the imaginary part of the mass vanishes and the real mass splits into two. The larger one has a sharp increase,
while the smaller one starts increasing when $\mu$ is large enough.

\section{$N$-point function}

In this Section, we consider the $N$-point function in the pure gauge theory ($\alpha=0$) in the confinement phase. Such function is related to the potential between $N$ static quarks (baryon potential). In order to compute the $N$-point function one should specify sources $\eta(x)$ and $\bar{\eta}(x)$ in Eq.(\ref{corr_1}). Let $x(i), i=1,\cdots,N$ be positions of $N$ static quarks on
a $d$-dimensional lattice. Then, one takes the following values: $\eta(x(i))=1$,
$\eta(x)=0$ if $x\ne x(i)$ and $\bar{\eta}(x)=0$ for all $x$. In the confinement region the saddle-point solutions equal that of the $U(N)$ model, Eq.(\ref{rho_0_un}). Expanding around these solutions, one finds
after a long but straightforward algebra
\begin{equation}
\Gamma_N(\sigma) \sim  \sum_x \ \prod_{i=1}^N \ G_{x,x(i)}(\sigma) \ , \
\sigma = \sqrt{\frac{2}{\beta}(1-d\beta)} \ ,
\label{Npoint_corr}
\end{equation}
where the sum over $x$ runs over all lattice sites and an irrelevant constant factor is omitted.  In the continuum limit the Green function (\ref{Green_func}) takes the form
\begin{equation}
G_{x,x^{\prime}} = \frac{\mbox{const}}{R^{\frac{d}{2}-1}} \ K_{\frac{d}{2}-1} (\sigma R) \ , \
R^2=\sum_{n=1}^d (x_n-x_n^{\prime})^2 \ ,
\label{green_func}
\end{equation}
where $K_n(x)$ is the modified Bessel function of the 2nd kind.
The summation over $x$ can now be replaced by the integration.
The evaluation of $\Gamma_N(\sigma)$ reduces to the well-known geometric median problem: find a point $y$ which minimizes the expression
\begin{equation}
\sum_{i=1}^N\sqrt{\sum_{n=1}^d(y_n-x_n(i))^2} \ .
\end{equation}
It follows, the $N$-quark potential takes the form of the geometric median law
\begin{equation}
V_N(\sigma) = - \ln\Gamma_N(\sigma) \sim  \sigma \ \sum_{i=1}^N |y-x(i)|  \ .
\label{GM_law}
\end{equation}
If $N=3$ this gives a $Y$ law for the three-quark potential
$V_3(\sigma) \ \sim \ \sigma Y$. This result agrees with Ref.~\cite{Ylaw_pl_models}, where it was shown via Monte-Carlo simulations
that the $Y$ law dominates the three-quark potential in $SU(3)$ Polyakov loop model. $\sigma$ is a string tension of the $N$-quark system. It equals the quark--anti-quark string tension. This elucidates how an analog of the $Y$ law appears in the large $N$ limit.

In order to calculate the $N$-point function above the phase transition one has to use Eqs.~(\ref{correlator_res})-(\ref{A2_notation}). One gets for the connected part the following expression
\begin{equation}
\Gamma_N(\sigma) \sim  M^N \sum_{i\ne j}
\left ( G_{x(i),x^{\prime}(j)}(m_1) - G_{x(i),x^{\prime}(j)}(m_2) \right ) \ .
\label{Npoint_corr_dec}
\end{equation}
This result leads to conclusions similar to those described in Sec.~\ref{compl_mass}.

\section{Summary}

This paper continues the investigation of Polyakov loop models at nonzero chemical potential in the 't Hooft-Veneziano limit. 
In \cite{pl_largeN21} we have studied the general phase structure of various such models. 
The present paper deals with the correlation functions of the Polyakov loops and the corresponding screening masses. Our main findings are the following:

\begin{itemize}
\item
Explicit formulas for the correlation functions and screening masses of $U(N)$ and $SU(N)$ Polyakov loop models with nonzero chemical potential have been obtained in the 't Hooft-Veneziano limit both in the confinement and deconfinement phases.

\item
In the deconfinement phase we established the existence of the complex masses and an oscillating decay of correlations in a certain region of parameters.

\item
The computation of the screening masses in different regions demonstrates that
at small $\alpha$ the ratio $m_i/m_r \gg 1$ reaches its maximum
close to the critical surface. In this region, one observes profound oscillations
of correlation functions with distance.

\item
It was shown, the calculation of the $N$-point correlation function reduces to the geometric median problem in the confinement phase.

\end{itemize}

In a nutshell, the paper provides a deeper understanding of the properties of the Polyakov loop models at finite density in the large $N$ limit.
It would be important and interesting to extend the results of the present paper
to models with an exact static determinant and to $SU(N)$ models at finite $N$.
Such work is in progress.

\section*{Acknowledgements}
V. Chelnokov acknowledges support by the Deutsche
Forschungsgemeinschaft (DFG, German Research Foundation) through 
the CRC-TR 211 ’Strong-interaction matter under extreme conditions’ – project number
315477589 – TRR 211. 
The work of S.~Voloshyn is supported by the National Academy of Sciences of Ukraine, Grant No. 0122U200259.

\section*{Appendix}

The functions $b_1, b_2, b_3$ are  given by
\begin{eqnarray}
&&b_1 = \frac{2 i t}{ s \Delta }   \bigg [ 4 t^4 (H_2+H_4)+4 H_4 s^2 t^2 +
i t s u (s+i t)^2   \nonumber   \\
\label{B_1}
&&+ \left (s^2+t^2 \right)^2 \left (H_3+u^2 \ln \frac{1 + u}{u} \right ) \bigg ]
+ \frac{2 i t^3}{s u}  \ , \\
\label{B_2}
&&b_2 = \frac{i u^2 }{2 \Delta }   \left[4 s t H_2  - i u (s + i t)^2\right] + \frac{1}{2}\alpha \rho \cosh \omega \ , \\
\label{B_3}
&&b_3 = \frac{2 i \, u t}{\Delta }  \left[ 4  t^2 H_2+ 2(s^2 + t^2) H_4 - u (s+ i t )^2   \right] -\alpha  \sin i \omega \ ,  \\
\label{Delta}
&&\Delta = 4 i  s t(H_2 H_3 - H_4^2) + u \left [ (s + i t)^2H_3  + 4 i s tH_4  - 4 t^2(H_2 + H_4) \right ]  \\
&&+ u^2 (4 i s t H_2 + u (s + i t)^2 )  \ln \frac{1 + u}{u} \ . \nonumber
\end{eqnarray}
The functions $H_2, H_3, H_4$ are defined as derivatives of the function $H_1$
\begin{align}
\label{Hi_def}
&H_1 = \sum_{k=0}^{\infty}  r^{2k+2} C_k(u) \ ,
H_2  = r^2 \frac{\partial^2 H_1}{\partial r^2 } \ , \
H_3 = u^2 \frac{\partial^2 H_1}{\partial u^2}   \ , \
H_4   = r u \frac{\partial H_1}{\partial u \partial r } \ , \\
\label{C_k}
&C_k(u) = \sum _{m=1}^{\infty} \frac{u^m(-4)^k (-1)^m \Gamma \left(k+\frac{m}{2}+1\right) \Gamma (2 k+m)}{(k+1) (k+1)! (2 k+1)! \Gamma \left(\frac{m}{2}+1\right) \Gamma (m)} \ , \ r=\sqrt{s^2+t^2} \ .
\end{align}
The saddle-point solutions near the critical surface $z\sim 0$ read \cite{pl_largeN21}
\begin{eqnarray}
\label{sun_saddles}
&&u_0 = y \ , \ t_0 = \frac{y}{2\rho_1} \ , \
s_0 = -i \left ( \frac{\alpha}{2} + \beta d\rho_1 +\beta d \rho_2 y \right ) \ ,
\omega_0 = -i \frac{y}{\alpha \rho_1} \ ,    \\
&&\rho_0 = \rho_1 + \rho_2 y \ , \
\rho_1 = \frac{\alpha}{2(1-\beta d)} \ , \
\rho_2 = \frac{1}{\alpha} \ \sqrt{1-4 \rho_1^2} \ , \
y = - \frac{z}{W_{-1} \left ( - e^{-c} z \right ) } \ ,  \nonumber
\end{eqnarray}
where $W_{-1}(x)$ is a lower branch of the Lambert function and $c$ -- unessential constant.


\begin{thebibliography}{99}


\bibitem{philipsen_rev_19}
O.~Philipsen, PoS LATTICE \textbf{2019} (2019) 273 [arXiv:1912.04827 [hep-lat]].
%
\bibitem{Seiler} E.~Seiler, EPJ Web Conf. \textbf{175}, (2018) 01019 [arXiv:1708.08254 [hep-lat]].
%
\bibitem{Gattringer11}
C.~Gattringer,
Nucl. Phys. B \textbf{850}, 242, (2011); [arXiv:1104.2503 [hep-lat]].
%
\bibitem{Gattringer12a}
Y.D.~Mercado, C.~Gattringer,
Nucl. Phys. B \textbf{862}, 737-750, (2012); [arXiv:1204.6074 [hep-lat]].
%
\bibitem{Philipsen12}
M.~Fromm, J.~Langelage, S.~Lottini, O.~Philipsen,
JHEP \textbf{01}, 042, (2012); [arXiv:1111.4953 [hep-lat]].
%
\bibitem{un_dual18} O.~Borisenko, V.~Chelnokov, S.~Voloshyn, EPJ Web Conf. {\bf 175} (2018) 11021 [arXiv:1712.03064 [hep-lat]].
%
\bibitem{pl_dual20} O.~Borisenko, V.~Chelnokov, S.~Voloshyn, Phys.Rev. D {\bf 102} (2020) 014502 [arXiv:2005.11073 [hep-lat]].
%
\bibitem{gross_witten}   D.~J.~Gross, E.~Witten, Phys.Rev. D {\bf 21} (1980) 446.
%
\bibitem{wadia} S.~R.~Wadia, Phys.Lett. B {\bf 93} (1980) 403.
%
\bibitem{damgaard_patkos} P.~H.~Damgaard and A.~Patk\'{o}s, Phys.Lett. B {\bf 172} (1986) 369.
%
\bibitem{christensen12} C.~H.~Christensen, Phys.Lett. B {\bf 714} (2012) 306
[arXiv:1204.2466 [hep-lat]].
%
\bibitem{largeN_sun} O.~Borisenko, V.~Chelnokov, S.~Voloshyn,
Nucl.Phys. B960 (2020) 115177 [arXiv:2008.00773 [hep-lat]].
%
\bibitem{pl_largeN_conf21} O.~Borisenko, V.~Chelnokov, S.~Voloshyn,  PoS LATTICE \textbf{2021} (2021) 453 [arXiv:2111.07103 [hep-lat]]
%
\bibitem{pl_largeN21} O.~Borisenko, V.~Chelnokov, S.~Voloshyn, Phys.Rev. D {\bf 105} (2022) 014501 [arXiv:2111.00474 [hep-lat]].
%
\bibitem{Bazavov} A.~Bazavov and J.H.~Weber, Progress in Particle and Nuclear Physics \textbf{116} (2021) 103823 [arXiv:2010.01873 [hep-lat]].
%
\bibitem{Nagata13}
J.~Takahashi, K.~Nagata, T.~Saito, A.~Nakamura, T.~Sasaki, H.~Kouno, M.~Yahiro,
Phys. Rev. D \textbf{88}, 114504 (2013) [arXiv:1308.2489 [hep-lat]].
%
\bibitem{Andreoli} M.~Andreoli, C.~Bonati, M.~D’Elia, M.~Mesiti, F.~Negro,
A.~Rucci, F.~Sanfilippo, Phys.Rev. D \textbf{97} (2018) 054515 [arXiv:1712.09996 [hep-lat]]
%
\bibitem{mcdual_21} O.~Borisenko, V.~Chelnokov, E.~Mendicelli, A.~Papa, Nucl.Phys.B \textbf{965} (2021) 115332 [arXiv:2011.08285 [hep-lat]].
%
\bibitem{su3_dual_cooreleations}
O.~Borisenko, V.~Chelnokov, E.~Mendicelli, A.~Papa,
\emph{Dual simulation of a Polyakov loop model at finite baryon
density: correlations and screening masses}, arXiv:2309.06104 [hep-lat].
%
\bibitem{Meisner2011}P. N.~Meisinger, M. C.~Ogilvie, T. D.~Wiser,
Int.J. Theor.Phys. \textbf{50} (2011) 1042 [arXiv:1009.0745 [hep-th]].
%
\bibitem{ogilvie2016} H.~Nishimura, M.~Ogilvie, K.~Pangeni,
Phys.Rev. D \textbf{93} (2016) 094501 [arXiv:1512.09131 [hep-lat]].
%
\bibitem{forcrand_z3} O.~Akerlund, P.~de Forcrand, T.~Rindlisbacher,
JHEP \textbf{10} (2016) 055 [arXiv:1602.02925 [hep-lat]].
%
\bibitem{duals_abelian}
O.~Borisenko, V.~Chelnokov, S.~Voloshyn, P.~Yefanov,
Phys.Lett. B {\bf 827} (2022) 137000
[arXiv:2112.06002 [hep-lat]].
%
\bibitem{Hooft_74} G.~t' Hooft, Nucl.Phys. B {\bf 72} (1974) 461.
%
\bibitem{Veneziano_76} G.~Veneziano, Nucl.Phys. B {\bf 117} (1976) 519.
%
\bibitem{Ylaw_pl_models} O.~Borisenko, V.~Chelnokov, E.~Mendicelli, A.~Papa,
Nucl.Phys.B {\bf 940} (2019) 214  [arXiv:1812.05384 [hep-lat]].



\end{thebibliography}
\end{document}